# On the functional properties of microRNA-mediated feed forward loops.

Yonatan Bilu, Department of Molecular Genetics, Weizmann Institute of Science, Rehovot 76100, Israel.

**Abstract**
**Motivation:** Recent studies of genomic-scale regulatory networks suggested that a feed-forward loop (FFL) circuitry is a key component of many such networks. This led to a study of the functional properties of different FFL types, where the regulatory elements are transcription factors.

**Results:** Here we investigate these properties when the mediating regulatory element of the loop is a microRNA. We find that many of the FFL properties are enhanced within this setup. We then proceed to identify all such FFLs in the D. Melanogaster regulatory network. We observe that in FFLs rooted at the same transcription factor there are significant correlations between the number of predicted binding sites for the transcription factor and for the microRNA.

**Conclusions:** Based on a modeling approach we suggest that these correlations may be an outcome of the type of FFL preferred by the transcription factor. This may help elucidate the type of regulation the TF confers.

**Background**

MicroRNAs (miRNAs) are short, highly conserved, endogenous RNA molecules (20-25bp long), that regulate protein expression by selectively binding to mRNA transcripts. Through recruitment of the RISC protein complex, they inhibit translation or initiate RNA cleavage. In recent years, the importance of miRNAs has become apparent, and it is estimated that they consist of at least 1% of the genes in Metazoa genomes, and that many of them target thousands of transcripts (reviewed, e.g. in (Bartel, 2004; Ke, et al., 2003)). However, despite their abundance and plausible importance, only little is known about their cellular roles.

Initially, miRNA were considered to have widespread essential roles, especially in cellular fate specification (see e.g. (Bartel, 2004; Ke, et al., 2003)). This view was supported by their tight conservation and the abundance of their targets. However, this perception was somewhat altered by several studies ((Giraldez, et al., 2005; Hornstein and Shomron, 2006; Stark, et al., 2005; Ying and Lin, 2005)). For example, Giraldez et al. (Giraldez, et al., 2005) showed that zebrafish embryos whose miRNA processing mechanism has been knocked out undergo surprisingly accurate axis formation and cell differentiation, leading them to suggest a modulating or tissue-specific role for miRNAs. Stark et al. (Stark, et al., 2005) showed that the expression of many miRNAs tends to be mutually exclusive with that of their target transcripts, suggesting a main role for mRNA in preventing translation of leaky basal transcription.

Hornstein and Shomron (Hornstein and Shomron, 2006) proposed that this modulating role is achieved by the wiring of miRNA within the cellular regulatory network. Specifically, they suggested that the expression of a miRNA will be regulated by a transcription factor (TF), which also jointly regulate with it the expression of a protein-coding gene, forming a feed forward loop. For example, if this TF activates the miRNA and the target-gene transcription , while the miRNA blocks the target-gene translation, the resulting feed forward loop (so called type-1 incoherent feed forward loop (Alon, 2007; Alonso and McKane, 2002); Figure 1) may confer robustness of the target-gene expression to variability in the TF levels (Alon, 2007).

Feed forward loops (FFLs) have been shown to be a major component of biological networks, and their functionality within the context of expression regulation has been analyzed both theoretically (Ghosh, et al., 2005; Mangan and Alon, 2003; Mangan, et al., 2003) and experimentally (Mangan, et al., 2006). These studies focused on FFLs in which both regulatory components are TFs, such that expression regulation is controlled at the transcriptional level. In this work we compare the functionality of such FFLs with the architecture suggested by Hornstein and Shomron, where the mediating regulator in the FFL is a miRNA (Figure 1). We find that such FFLs generally facilitate previously suggested functions more readily. However, this facilitation comes at the cost of higher noise level of the protein product.

To examine the properties of such miRNA-FFLs in a real network, we construct a large-scale sequence-based enumeration of them in Drosophila melanogaster. We investigate the strength of the regulatory connections in FFLs, a parameter that can be deduced from sequence analysis or from DNA binding data, but so far has received little attention in this context. We detect a puzzling correlation between the number of TF binding sites in a miRNA promoter, and the number of binding sites for the miRNA in the 3' UTR of its predicted target. We suggest that this correlation may be the outcome of different types of FFLs. This, in turn, may shed light on the type of regulatory connection (positive or negative), as these can be deduced from the type of FFL in which they are involved.

# Results

*Comparison of miRNA-mediated and TF-mediated FFLs*

We start by comparing the functional properties of miRNA-mediated and TF-mediated FFLs. The general setting for this analysis is shown in Figure 1. The FFL is rooted at a TF (X) which responds to an outside signal. When active, it activates or inhibits the expression of both a downstream protein-coding gene (Z), and an additional regulatory element which may be a TF or a miRNA (Y; Figure 1). Importantly, the protein-coding target gene is also regulated by this additional regulatory element.

The main interest in studying FFLs is the behavior and dynamics of the level of protein-coding gene target Z, which may be thought of as the "output" of the FFL. We used a modeling approach to examine three properties of FFLs: (i) "pulse generation" (Mangan and Alon, 2003) – a rapid rise in the level of Z, which is then followed by a gradual decrease to the steady state level; (ii) "response time" (Mangan and Alon, 2003) – the time it takes for Z to reach half its steady state level (or initial level, if the latter is zero). FFLs are known to either quicken or delay the response, depending on the specific architecture (Mangan and Alon, 2003); and (iii) "noise" – the coefficient of variance for the steady-state level of Z.

Our modeling approach and choice of values for the reaction constants roughly follows that of Mangan and Alon (Mangan and Alon, 2003), and is depicted for the type-1 incoherent FFL in Figure 2 (also, see Methods). Figure 3 shows the dynamics of the level of Z resulting from these reaction schemes. Modeling of the type-3 and type-4 coherent FFLs is done similarly, and is described in the Supplementary Data.

When comparing the properties of the two types of FFLs, we observed several possible advantages of miRNA-FFLs over TF-FFLs (Table 1; see Methods). First, miRNA-FFLs are RNA-based, rather than protein-based, and hence faster and more cost-effective (require less energy). Second, the pulse generation of type-1 incoherent FFLs is stronger in miRNA-FFLs. Third, the response time of both type-1 incoherent FFLs and type-3 coherent FFLs, are shorter in miRNA-FFLs. Finally, the delayed response time (Mangan and Alon, 2003) of type-4 coherent FFLs is more effective in miRNA-FFLs.

On the other hand, as shown in Table 1, miR-FFLs have a much noisier output. Using the Gillespie algorithm ((Gillespie, 2006); implemented in Dizzy (Ramsey, et al., 2005)), we simulated the level of the protein Z at steady state. We found that this level is noisier in the miRNA-FFL than in the TF-mediated one, especially in the type-4 incoherent FFL and the type-3 coherent FFL.

Taken together, these results may support the conjecture of Hornstein and Shomron (Hornstein and Shomron, 2006), that a likely architecture for miRNA regulation within the global network is a FFL. However, it suggests that a TF-based and a miR-based FFL may have different properties and are each optimal for different scenarios. While a TF-based FFL is more appropriate for noise reduction, the use of a miRNA component enhances other functionalities of the FFL, including pulse generation, response time and delayed response.

Under plausible simplifying assumptions, the differential equations associated with the reaction schemes can be solved analytically, as described in the Supplementary Data.

*Identifying miRNA-FFLs in the D. melanogaster regulatory network*
The basic architecture of a FFL is as in Figure 1. However, this schematic depiction does not take into account the strength or cooperativity of the regulatory connections. To study this aspect, we compiled a genomic-scale list of miRNA-FFLs in D. melanogaster. We specifically focused on the number of TF and miRNA binding sites, which represents the strength of the regulatory connection.

To this end, we first reconstructed the regulatory connections among TFs, miRNAs and protein-coding genes (see Methods). Briefly, this construction is based on all known 14226 protein-coding genes (downloaded from FlyBase (Crosby, et al., 2007)), the 78 known miRNAs (downloaded from miRBase (Griffiths-Jones, et al., 2006)), and 371 TF binding site motifs identified by Elemento and Tavazoie (Elemento and Tavazoie, 2005) (which we further clustered down to 321). Regulation by miRNAs was taken from miRBase; regulation by TFs was determined by searching for cis-regulatory modules (CRMs) encompassing clusters of TF binding sites within proximal promoters of protein-coding and miRNA genes (see Methods).

A similar analysis was also performed for H. sapiens, based on the TF binding site motifs identified by Xie et al. (Xie, et al., 2005). The construction and resulting network will be discussed elsewhere (submitted), and here we focus on the strength of the regulatory connections within identified miR-FFLs.

Having identified all these regulatory connections, we obtained for each TF the list of miRNA-FFLs it is involved in. This is simply all miRNA-gene pairs, such that the miRNA targets the gene, and the TF has a binding site (within a CRM) in the promoter region of both the miRNA and the gene. All in all, we identified 3945 such FFLs. In addition, for each edge in the FFL, we noted its "strength" – the number of TF (or miRNA) binding sites within the promoter (or 3'-UTR).

*Correlated number of binding sites in FFLs*

We next examined the correlation between the number of TF binding sites in the miRNA promoter, and the number of binding sites for that miRNA in the 3' UTR of the downstream target genes. Formally, for a TF T, involved in k FFLs, we constructed two vectors, $v_T^{TF->miR}$ and $v_T^{miR->gene}$, of dimension k. The ith entry of $v_T^{TF->miR}$ is the number of binding sites for T in the promoter region of the miRNA of the ith FFL, and the ith entry of $v_T^{miR->gene}$ is the number of binding sites for that miRNA in the 3' UTR of the downstream target gene (Figure 4a). We then computed the Spearman rank correlation between $v_T^{TF->miR}$ and $v_T^{miR->gene}$, and the corresponding p-value.

Of the 321 TFs, 189 participate in at most one FFL, and for an additional 98 one of the vector $v_T^{TF->miR}$ or $v_T^{miR->gene}$ is constant. In all these cases, the Spearman rank correlation between the vectors is undefined. However, as shown in Figure 4b, the remaining 34 TFs display a puzzling bi-modal distribution of correlation coefficients (Lilliefors normality test: α=0.01, p-value < 0.01), 16 of which are associated with a highly significant p-value (p-value < 10-3; false discovery rate = 0.0021). Moreover, this bi-model distribution and high number of significant correlations is also evident for FFLs derived from the H. Sapiens network (13 of 21 have p-value < 10-3). Interestingly, no significant correlation was found between the number of TF binding sites in the promoter region of the downstream gene and either $v_T^{TF->miR}$ or $v_T^{miR->gene}$.

What might be the reason for this bi-modality, and the seemingly significant correlations? A possible explanation is provided by considering the effect of multiple binding sites in the modeling procedure described above. Specifically, we compared the functional properties of the miRNA-FFLs when substituting the reaction $X + P_y \leftrightarrow [XP_y]$ by $2X + P_y \leftrightarrow [2XP_y]$, and the reaction $Y + Z \rightarrow [YZ] \rightarrow Y$ by $2Y + Z \rightarrow [2YZ] \rightarrow 2Y$.

Focusing on the example of the type-1 incoherent feed forward loop, our modeling suggests that two binding sites in the 3' UTR increases the strength of the pulse generated by the miRNA-FFL and shortens its response time. In contrast, two TF binding sites in the miRNA's promoter have little effect on functionality (Table 2), and thus one of them is likely to accumulate mutations and degenerate. This may imply that type-1 incoherent FFLs have a preference towards a specific pattern of binding sites – one TF binding site in the miRNA's promoter, and multiple miRNA binding sites in the gene's 3' UTR (i.e. $v_T^{TF->miR}=1$ and $v_T^{miR->gene}>1$).

Such a preference will lead to a seemingly significant anti-correlation between $v_T^{TF->miR}$ and $v_T^{miR->gene}$, over a set of type-1 incoherent FFLs. Intuitively, this is clear; in many FFLs the number of binding sites in the miRNA promoter will be one, while their number in the gene's 3' UTR will be high. More formally, suppose that among a set of 50 FFLs (which is indeed the average number of FFLs per TF in our data) 70% display the preferred pattern, while each of the other possible patterns is displayed in 10% of the FFLs. In this case, the Spearman rank correlation is,-0.375 (p=0.007). Furthermore, by generating such sets at random, we estimated that with probability 0.35, the p-value is less than 10-3.

The same line of reasoning suggests that in sets of type-3 coherent FFLs a similar negative correlation will tend to appear, while in sets of type-4 coherent FFLs a positive correlation will tend to appear (see Supplementary Data). Taken together, this may account for both the seemingly significant correlations among binding sites numbers, and their bi-modal distribution.

As an example, we looked for TF motifs from (Elemento and Tavazoie, 2005) which match (according to TransFac (Matys, et al., 2006)) the recognition site of the TF Ftz, a known activator, or of Snail, known to act both as an activator and inhibitor. Indeed, we find that Ftz is associated with 3 motifs with a defined correlation between $v_T^{TF->miR}$ and $v_T^{miR->gene}$, all three of which are negative, as is expected in a type-1 incoherent FFL, where the TF acts solely as an activator. Similarly, we find that the 9 motifs associated with Snail display both negative (5 motifs) and positive (4 motifs) correlations, compatible with the behavior suggested for the incoherent FFLs, where the TF acts as activator of one element, and an inhibitor of the other.

An implicit assumption here is that FFLs rooted at the same TF will tend to be of the same type. This seems plausible according to data from E. Coli compiled by Shen-Orr et al. (Shen-Orr, et al., 2002), where all five TFs that are involved in multiple FFLs, display a preference for one type of FFL and regulatory connection (negative or positive), with which the majority of the FFLs comply (Table 3).

**Discussion and Conclusions**

In this work we analyzed the functional properties of FFLs in which the mediating regulatory element is a miRNA. We have shown that although the reaction scheme for such FFLs is very similar to that of TF-mediated ones, the dynamics of the downstream protein product is very different. Our analysis also supports a previously suggested conjecture that miRNAs tend to be involved in FFLs (Hornstein and Shomron, 2006). However, the arising explanation is different: miRNA FFLs enhance the functionality of the FFL rather than reduce noise at the protein level.

Studying the strength of the regulatory connections, we observed that in FFLs rooted at the same TF there tends to be a seemingly significant positive or negative correlation between the number of binding site in the miRNA promoter region and the number of binding sites for the miRNA in the 3' UTR of the joint target gene. A possible explanation for this is suggested by modeling the effect of multiple binding sites on the FFL functionality: Different coherency and architecture of FFLs may dictate a preferred arrangement of binding site numbers, leading to these seemingly significant correlations, as well as to the observed bi-model distribution of correlation coefficients.

The contribution of the latter observation is two-fold. First, we explore the impact of multiple binding sites on FFL functionality, highlighting certain binding-site arrangements as more likely for a given type of FFL. Second, this analysis suggests that even in sequence-based constructions of regulatory network, it is possible, in specific cases, to predict negative or positive regulation. Namely, when a regulatory connection is part of a FFL, whose specific type is suggested by analyzing the correlation between binding site numbers. For example, TFs involved mainly in type-1 incoherent FFLs would be predicted to confer mainly positive regulation.

Notably, a broader integration of other network properties are required for a refined characterization of regulatory interactions. Differentiating between positive and negative correlations is not sufficient, as this distinction only partitions the FFLs into two sets. Identifying the exact type of FFL is required in order to elucidate the type of all FFL-integrated regulatory connections. Future work will hopefully incorporate properties such as the value of the correlation coefficient, or the actual number of binding sites, to achieve this goal.

**Methods**

*Datasets used*

Genomic sequences of D. Melanogaster were download from FlyBase (version 4.2; http://flybase.bio.indiana.edu/), as well as CDs and locations of protein-coding genes.

D. Melanogaster miRNAs and their targets were downloaded from miRBase ((Griffiths-Jones, et al., 2006); http://microrna.sanger.ac.uk/), which lists 78 miRNAs (23 intronic, 55 intergenic), with an average of 89.56 targets for each miRNA.

TF binding sites motifs were taken from the list compiled by Elemento and Tavazoie (Elemento and Tavazoie, 2005) for the D. Melanogaster network (371 highest scoring k-mers), and from the list of 198 motifs of Xie et al. (Xie, et al., 2005) for the H. Sapiens one.

DNA footprinting data was taken from the Drosophila DNase I Footprint Database (v2.0; (Bergman, et al., 2005); http://www.flyreg.org/), which provides data for 87 TFs and 101 target genes, for a total of 1365 "footprints".

Upstream sequences of human miRNAs and protein coding genes (2000bp) were downloaded from the UCSC Genome Browser (http://genome.ucsc.edu/).

*Promoter region definition*

For protein-coding genes, the 2000bp upstream of the transcription start site were defined as the promoter region of the gene (as in (Elemento and Tavazoie, 2005)).

For intronic miRNAs, we assumed that miRNAs are expressed together with their host genes (Baskerville and Bartel, 2005), and defined their promoter region to be that of the host genes.

For intergenic miRNAs, the 2000bp upstream of the identified pre-miRNA are defined as the promoter region of the gene. In this case, we also clustered polycistronic miRNAs, such that two miRNAs that are at most 500bp apart are in the same cluster. The promoter of a polycistronic cluster starts from the first miRNA in the cluster.

Defining promoter regions of intergenic miRNAs this way seems plausible for two reasons. First, these regions are as enriched for CRM motif appearances as the promoter regions of protein coding genes (9.75 sites on average in miRNA promoters, and 9.00 in protein-coding genes promoters). Second, they are distinctly more enriched than random sequences with the same di-nucleotide distribution (Z-score = 19.52). Notably, these two observations also support the implicit assumption that miRNAs are regulated by the same TF as protein coding genes (indeed, it is known the miRNAs are transcribed by RNA Polymerase II (Baskerville and Bartel, 2005)).

*Motif detection*

Identifying functional TF recognition motifs within promoter sequences is a challenging task, since many spurious appearances are expected simply by chance. For example, a motif of average length in our dataset (7bp) is expected to appear approximately once in every four promoter regions by chance alone ($4^7/(2*2000) = \sim 4$). To overcome this, we examined three methodologies:

Near TSS (transcription start site): All motif matches within 300bps upstream of the gene were considered genuine. This approach is motivated by several independent observations (Harbison, et al., 2004; McCue, et al., 2001; Xie, et al., 2005) that this short region contains most of the functional motif appearances. Considering only short sequences (300bp), decreases the probability of falsely identifying spurious appearances by about one order of magnitude. However, as the transcription start site of miRNAs is unknown, we note that this approach will often result in DNA fragments that are embedded within the transcribed pri-miRNA precursor form.

Conservation: First, all motifs within the promoter region were identified. Then, for each identified motif the aligned sequences in the seven other Drosophila species (D. simulans, D. yakuba, [D. erecta](), D. ananassae, D. pseudoobscura, D. mojavensis and D. virilis) were analyzed. If in at least 5 of them the motif appeared with at most 1 mismatch, the occurrence was taken as genuine. This approach is the one most commonly used in identifying true motif appearances (Grun, et al., 2005; Xie, et al., 2005), and is motivated by the plausible assumption that spurious appearances will tend to mutate during evolution, while functional ones will tend to be conserved.

(CRMs): First, all motifs within the promoter region were identified. Then, for each identified motif a 99bps window centered at it was taken, and the number of identified motifs within this window was computed. If this number was at least 10, the motif occurrence was taken to be genuine. This methodology formalizes the observation that TF binding sites in Drosophila tend to be arranged in clusters ((Levine and Davidson, 2005)). Adding this constraint helps to eliminate spurious appearances, which are expected to be uniformly distributed in a 2000bp region, and not form clusters.

The CRMs method proved best in identifying experimentally determined TF binding site (via DNA footprints, (Bergman, et al., 2005); Supplementary Table 1), and was thus used for the analysis described in the text, both for D. Melanogaster and H. Sapiens. We believe that sequence conservation would have been an equally powerful approach if the study was not conducted with closely related species. The evolutionary time frame for the divergence of genomes within Drosophila species may be too short to allow for detection of only functionally 'conserved' sequences. Furthermore, since CRMs is a measure independent of conservation it provides a tool for detection of TF binding sequences that is independent from the methodology by which the TF motifs were originally defined (Elemento and Tavazoie, 2005).

*Clustering of motifs*

Some of the motifs identified by Elemento and Tavazoie (Elemento and Tavazoie, 2005) are highly similar. After constructing the network, such motifs were clustered together, and the number of "hits" for such a cluster was taken as the maximal among its constituents.

Thus, for each pair of motifs we computed a similarity score using the smith-waterman algorithm, with 1 for a match and 0 for a mismatch or a gap, and normalized it by the motif's length. We then constructed a graph of these motifs, with adjacency defined by a similarity score of at least 0.8. For each connected component in this graph, we iteratively threw out the motif which had the least average similarity to the rest of the motifs in the component, until the similarity between any two motifs was at least 0.7. These remaining motifs were taken as a cluster.

This clustering algorithm reduced the number of represented TFs from 371 to 321. However, we found that this clustering had a negligible effect on our analysis - essentially identical results were obtained on the un-clustered data.

*Modeling of miRNA -FFLs*

The kinetic constants for the reaction schemes were taken to be 1 for TF-promoter dissociation constants, degradation constants and species formation rates (except for miRNA formation, which is taken to be 2 when so noted). The basal production rate of Y was taken to be 0.5 and of Z to be 0 (as in Mangan and Alon (Mangan and Alon, 2003)).

The initial level of the chemical species is 100 for X, 1 for the promoters and 0 for all others, except in the type-3 coherent FFL, where the initial level of Z is 100, and in the type-4 coherent FFL, where the initial level of Y is 10.

Pulse strength and response time (time to reach half final/initial Z level) were computed based on numerical solutions to the differential equations. CV was computed by first numerically identifying the time point from which on Z's level is essentially constant, and then using the Gillespie algorithm to simulate the end product level from this time point on. All other functional parameters were found numerically. Both computations were done with the help of the Dizzy software (Ramsey, et al., 2005).

*Simulation of Spearman correlations*

We considered four types of binding sites arrangements in the miRNA promoter and the targeted mRNA's 3' UTR: (i) one binding site in each; (ii) two binding sites in each; (iii) one binding site in the former and two in the latter; (iv) one binding site in the latter and two in the former.

Next, given a distribution over these four types arrangements, we generated at random 1000 pairs of random vectors in $\{1,2\}^{50}$ by choosing each coordinate independently according to the given distribution, and for each pair calculated the Spearman rank correlation and p-value.

For example, consider the distribution of choosing arrangement (iii) with probability 0.7, and each of the others with probability 0.1. Then for every i=1,…,50; with probability 0.7 the ith entry is 1 in the first vector and 2 in the second; with probability 0.1 it is 1 in both; with probability 0.1 it is 2 in both; with probability 0.1 it is 2 in the first vector and 1 in the second.

The number 50 was chosen since it is approximately the average number FFLs per TF for which the Spearman correlation is well defined (in the Drosophila network).

**Acknowledgements**

We thank Uri Alon, Eran Hornstein, Ilya Soifer and Noa Rappaport for their advice and comments. This work was partially funded by the UniNet EC NEST consortium contract number 12990 to YB and a Minerva grant and a JDRF grant to EH.

**References**

Alon, U. (2007) Network Motifs in Transcription Regulation Networks, *Nature Reviews Genetics*.

Alonso, D. and McKane, A. (2002) Extinction dynamics in mainland-island metapopulations: an N-patch stochastic model, *Bull Math Biol*, **64**, 913-958.

Bartel, D.P. (2004) MicroRNAs: genomics, biogenesis, mechanism, and function, *Cell*, **116**, 281-297.

Baskerville, S. and Bartel, D.P. (2005) Microarray profiling of microRNAs reveals frequent coexpression with neighboring miRNAs and host genes, *Rna*, **11**, 241-247.

Bergman, C.M., Carlson, J.W. and Celniker, S.E. (2005) Drosophila DNase I footprint database: a systematic genome annotation of transcription factor binding sites in the fruitfly, Drosophila melanogaster, *Bioinformatics*, **21**, 1747-1749.

Crosby, M.A., Goodman, J.L., Strelets, V.B., Zhang, P. and Gelbart, W.M. (2007) FlyBase: genomes by the dozen, *Nucleic Acids Res*, **35**, D486-491.

Elemento, O. and Tavazoie, S. (2005) Fast and systematic genome-wide discovery of conserved regulatory elements using a non-alignment based approach, *Genome Biol*, **6**, R18.

Ghosh, B., Karmakar, R. and Bose, I. (2005) Noise characteristics of feed forward loops, *Phys Biol*, **2**, 36-45.

Gillespie, D.T. (2006) Stochastic Simulation of Chemical Kinetics, *Annu Rev Phys Chem*.

Giraldez, A.J., Cinalli, R.M., Glasner, M.E., Enright, A.J., Thomson, J.M., Baskerville, S., Hammond, S.M., Bartel, D.P. and Schier, A.F. (2005) MicroRNAs regulate brain morphogenesis in zebrafish, *Science*, **308**, 833-838.

Griffiths-Jones, S., Grocock, R.J., van Dongen, S., Bateman, A. and Enright, A.J. (2006) miRBase: microRNA sequences, targets and gene nomenclature, *Nucleic Acids Res*, **34**, D140-144.

Grun, D., Wang, Y.L., Langenberger, D., Gunsalus, K.C. and Rajewsky, N. (2005) microRNA target predictions across seven Drosophila species and comparison to mammalian targets, *PLoS Comput Biol*, **1**, e13.

Harbison, C.T., Gordon, D.B., Lee, T.I., Rinaldi, N.J., Macisaac, K.D., Danford, T.W., Hannett, N.M., Tagne, J.B., Reynolds, D.B., Yoo, J., Jennings, E.G., Zeitlinger, J., Pokholok, D.K., Kellis, M., Rolfe, P.A., Takusagawa, K.T., Lander, E.S., Gifford, D.K., Fraenkel, E. and Young, R.A. (2004) Transcriptional regulatory code of a eukaryotic genome, *Nature*, **431**, 99-104.

Hornstein, E. and Shomron, N. (2006) Canalization of development by microRNAs, *Nat Genet*, **38 Suppl**, S20-24.

Ke, X.S., Liu, C.M., Liu, D.P. and Liang, C.C. (2003) MicroRNAs: key participants in gene regulatory networks, *Curr Opin Chem Biol*, **7**, 516-523.

Levine, M. and Davidson, E.H. (2005) Gene regulatory networks for development, *Proc Natl Acad Sci U S A*, **102**, 4936-4942.

Mangan, S. and Alon, U. (2003) Structure and function of the feed-forward loop network motif, *Proc Natl Acad Sci U S A*, **100**, 11980-11985.

Mangan, S., Itzkovitz, S., Zaslaver, A. and Alon, U. (2006) The incoherent feed-forward loop accelerates the response-time of the gal system of Escherichia coli, *J Mol Biol*, **356**, 1073-1081.

Mangan, S., Zaslaver, A. and Alon, U. (2003) The coherent feedforward loop serves as a sign-sensitive delay element in transcription networks, *J Mol Biol*, **334**, 197-204.

Matys, V., Kel-Margoulis, O.V., Fricke, E., Liebich, I., Land, S., Barre-Dirrie, A., Reuter, I., Chekmenev, D., Krull, M., Hornischer, K., Voss, N., Stegmaier, P., Lewicki-Potapov, B., Saxel, H., Kel, A.E. and Wingender, E. (2006) TRANSFAC and its module TRANSCompel: transcriptional gene regulation in eukaryotes, *Nucleic Acids Res*, **34**, D108-110.

McCue, L., Thompson, W., Carmack, C., Ryan, M.P., Liu, J.S., Derbyshire, V. and Lawrence, C.E. (2001) Phylogenetic footprinting of transcription factor binding sites in proteobacterial genomes, *Nucleic Acids Res*, **29**, 774-782.

Ramsey, S., Orrell, D. and Bolouri, H. (2005) Dizzy: stochastic simulation of large-scale genetic regulatory networks, *J Bioinform Comput Biol*, **3**, 415-436.

Shen-Orr, S.S., Milo, R., Mangan, S. and Alon, U. (2002) Network motifs in the transcriptional regulation network of Escherichia coli, *Nat Genet*, **31**, 64-68.

Stark, A., Brennecke, J., Bushati, N., Russell, R.B. and Cohen, S.M. (2005) Animal MicroRNAs confer robustness to gene expression and have a significant impact on 3'UTR evolution, *Cell*, **123**, 1133-1146.

Xie, X., Lu, J., Kulbokas, E.J., Golub, T.R., Mootha, V., Lindblad-Toh, K., Lander, E.S. and Kellis, M. (2005) Systematic discovery of regulatory motifs in human promoters and 3' UTRs by comparison of several mammals, *Nature*, **434**, 338-345.
Ying, S.Y. and Lin, S.L. (2005) MicroRNA: fine-tunes the function of genes in zebrafish, *Biochem Biophys Res Commun*, **335**, 1-4.

**Tables:**

|  | Response time | | | Pulse | CV | | |
|---|---|---|---|---|---|---|---|
|  | I-1 | C-3 | C-4 | I-1 | I-1 | C-3 | C-4 |
| TF-FFL | 13 | 14 | 14 | 1 | 1.07 | 1.39 | 0.95 |
| miRNA-FFL | 6 | 10 | 46 | 1.2 | 1.56 | 2.12 | 1 |
| miRNA-FFL, quick maturation | 5 | 10 | 46 | 1.4 | 1.85 | 2.40 | 1 |

**Table 1:** Values for functionally-relevant properties of FFLs. *Response time* is the time when the end product (Z) reaches half its final level (I-1, C-4) or half its starting level (C-3). *CV* is the ratio between the standard deviation of Z's level and its mean value, at steady state. *Pulse* is the ratio between the highest value attained by Z and its steady state one. *I-1* Type-1 incoherent FFL; *C-3* Type-3 coherent FFL; *C-4* Type-4 coherent FFL.

|  | One BS in 3' UTR | | | Two BSs in 3' UTR | | |
|---|---|---|---|---|---|---|
| Number of BSs in miRNA promoter | Pulse | CV | Resp. time | Pulse | CV | Resp. time |
| 1 | 1.4 | 1.85±0.15 | 5 | 2.72 | 2.01±0.23 | 3 |
| 2 | 1.38 | 1.85±0.16 | 4 | 2.73 | 1.91±0.21 | 3 |

**Table 2:** The effect of cooperativity on the values of the functional properties of the type-1 incoherent miRNA-FFL. Properties are is in Table 1 (last line).

| TF | Main FFL type | Number of FFLs of main type | Total number of FFLs |
|---|---|---|---|
| crp | type-1 coherent | 11 | 16 |
| flhDC | type-1 coherent | 5 | 5 |
| fnr | type-3 coherent | 4 | 7 |
| rob | type-1 coherent | 4 | 4 |
| rpoN | type-1 coherent | 4 | 4 |

**Table 3:** Distribution of FFL types according to the TF at which they are rooted. The main FFL type is the FFL type most commonly associated with the TF. The third column lists the number of FFLs of this type associated with the TF, and the fourth column the total number of FFLS associated with it.

The fraction of type-1 coherent FFLs in the network is 0.67; The fraction of type-3 coherent FFLs in the network is 0.1.

**Figures**

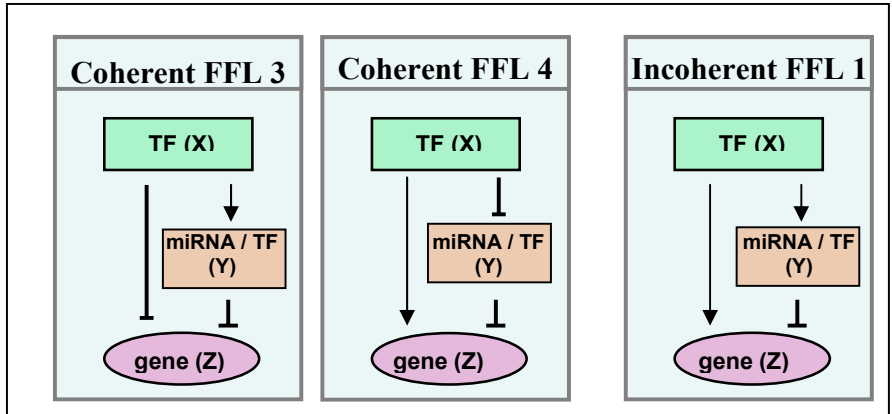

**Figure 1:** The architecture of a TF or miRNA mediated FFLs of the three types that relevant for miRNA regulation.

| | | |
|---|---|---|
| (1) $X + Py \leftrightarrow [XPy]$ | $X$ binds $Y$'s promoter and activates it. | **TF-mediated FFL:** $Y + Pz \leftrightarrow [YPz]$ $Y$ blocks $Z$'s promoter. |
| (2) $[XPy] \to [XPy] + Y$ | $Y$ is transcribed from its active promoter. | |
| (3) $X + Pz \leftrightarrow [XPz]$ | $X$ binds $Z$'s promoter and activates it. | |
| (4) $[XPz] \to [XPz] + Z$ | $Z$ is transcribed from its active promoter. | **miRNA-mediated FFL:** $Y + Z \to [YZ] \to Y$ $Y$ blocks $Z$'s mRNA from translation. Eventually $Z$ is degraded. |
| (5) $\to Y \to \phi$ | $Y$'s basal transcription and degradation. | |
| (6) $\to Z \to \phi$ | $Z$'s basal transcription and degradation. | |

**Figure 2:** Reaction scheme for TF and miRNA mediated FFLs. Reactions listed on the left side are relevant for both TF-FFLs and miRNA-FFLs. To the right are the reactions specific to each of the architectures.

X is a TF encoding the inducing signal. Z is the mRNA product of the circuit. Y is the inhibiting TF in the TF-mediated FFL, and the miRNA in the miRNA-mediated one. Pz and Py are the promoter regions of Z and Y. Species in square brackets are complexes of either TF-bound DNA ([XPy], [XPz], [YPz]) or miRNA-mRNA ([YZ]).

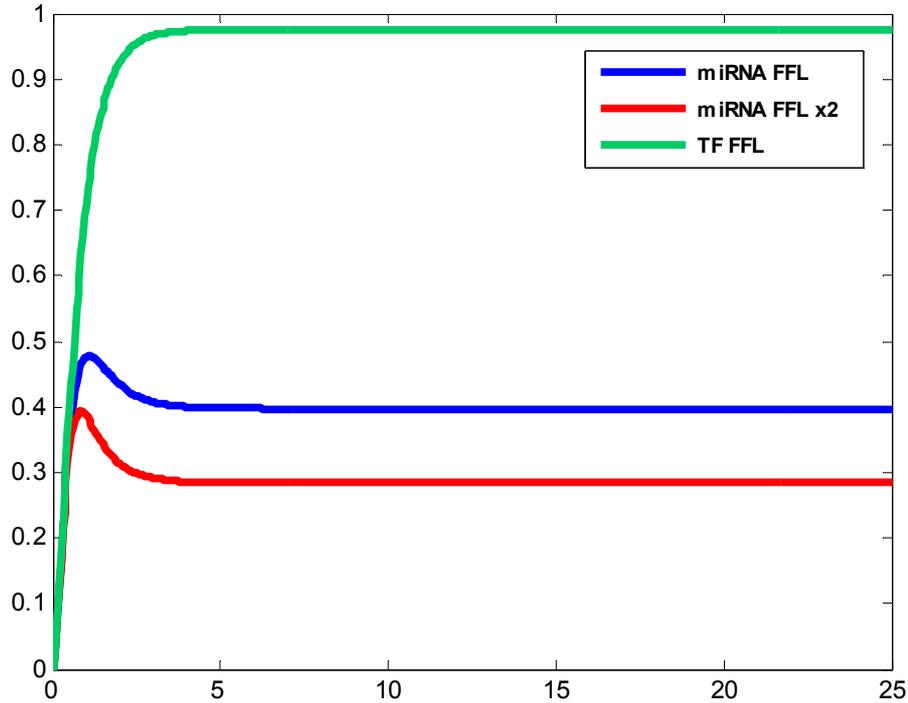

**Figure 3:** Level of the protein coding gene Z in the TF-FFL (green), in the miRNA-FFL (blue), and (red) in the miRNA-FFL, where the reaction constant for miRNA formation (reaction (2) in Figure 2) is taken to be twice as fast.

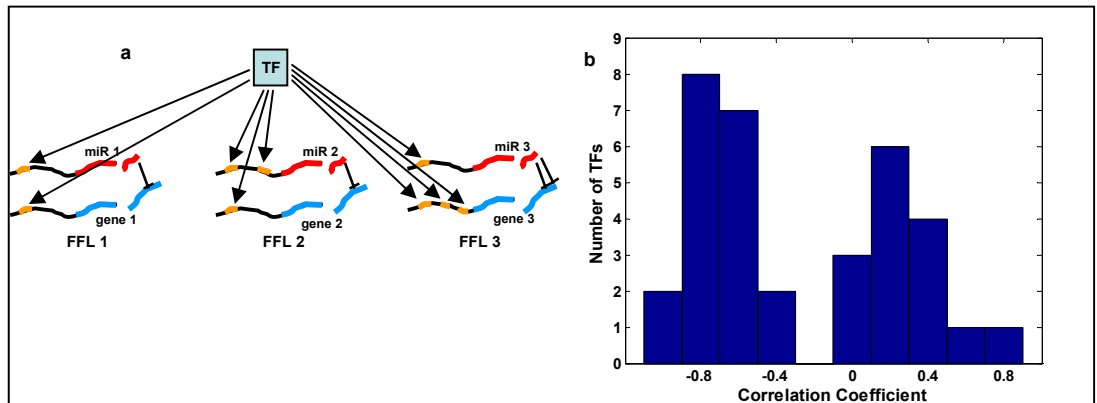

**Figure 4:** (*a*) A toy example for the construction of $v_T^{TF->miR}$ and $v_T^{miR->gene}$: here the TF has one binding site in the promoter region of the miRNA in the first FFL, two binding sites in the second, and one in the third, so $v_T^{TF->miR} = (1,2,1)$. Similarly, the miRNA in the first FFL has one binding site in the 3' UTR of the joint target gene, that in the second FFL also has one, and that in third FFL has two. Hence, $v_T^{miR->gene} = (1,1,2)$. (*b*) Distribution of correlation coefficients between $v_T^{TF->miR}$ and $v_T^{miR->gene}$ over 34 TFs.